\documentclass[pdflatex,sn-mathphys]{sn-jnl}

\usepackage{subfig}
\usepackage{natbib}

\jyear{2022}%

\theoremstyle{thmstyleone}%
\theoremstyle{thmstyletwo}%

\theoremstyle{thmstylethree}%

\newcommand\revision[1]{\textcolor{black}{#1}}

\begin{document}

\title[From Curious Hashtags to Polarized Effect]{From Curious Hashtags to Polarized Effect: Profiling Coordinated Actions in Indonesian Twitter Discourse}


\author[1]{\fnm{Adya} \sur{Danaditya}\textsuperscript{*}}\email{adanadit@andrew.cmu.edu}

\author[1]{\fnm{Lynnette} \sur{Hui Xian Ng}\textsuperscript{*}}\email{lynnetteng@cmu.edu}

\author[1]{\fnm{Kathleen} \sur{M. Carley}}\email{kathleen.carley@cs.cmu.edu}

\affil[1]{\orgdiv{Center for Computational Analysis of Social and Organizational Systems}, \orgname{Carnegie Mellon University}, \orgaddress{\street{4665 Forbes Avenue}, \city{Pittsburgh}, \postcode{15213}, \state{PA}, \country{USA}}}
\affil[*]{Equal Contributions}

\abstract{Coordinated campaigns in the digital realm have become an increasingly important area of study due to their potential to cause political polarization and threats to security through real-world protests and riots.
In this paper, we introduce a methodology to profile two case studies of coordinated actions in Indonesian Twitter discourse. 
Combining network and narrative analysis techniques, this six-step pipeline begins with DISCOVERY of coordinated actions through hashtag-hijacking; identifying WHO are involved through the extraction of discovered agents; framing of what these actors did (DID WHAT) in terms of information manipulation maneuvers; TO WHOM these actions were targeted through correlation analysis; understanding WHY through narrative analysis and description of IMPACT through analysis of the observed conversation polarization.
We describe two case studies, one international and one regional, in the Indonesian Twittersphere. 
Through these case studies, we unearth two seemingly related coordinated activities, discovered by deviating hashtags that do not fit the discourse, characterize the coordinated group profile and interaction, and describe the impact of their activity on the online conversation.
}

\keywords{information operations, coordination detection, narrative analysis, social network analysis, coordinated group identification, coordinated group profiling}

\maketitle

\section{Introduction}
The emergence of coordinated activity on the Web, especially social media, where groups of accounts work together to manipulate conversations as part of religious and political activism, has been a subject of increasing consequence in the last decade \cite{DBLP:journals/corr/abs-2107-02588}. Inorganic coordination between communities threaten society's social fabric, with the potential to evolve into harmful societal violence, e.g. the coordinated effort to propagate extremist ideas by ISIS \cite{awan2017cyber,zahrah2020isis}. 

Studies of coordinated activity have been performed in many regions of the world, including United States \cite{10.1145/3501247.3531542,DBLP:journals/corr/abs-2105-07454}, United Kingdom \cite{Nizzoli_Tardelli_Avvenuti_Cresci_Tesconi_2021} and Russia \cite{DBLP:journals/corr/abs-2107-02588,https://doi.org/10.48550/arxiv.2206.03576}.  
Meanwhile, little work has been done in the context of Southeast Asia, a region of increasing importance \revision{that houses} a diverse mix of 8.5\% of the world's population \cite{MR-1170-AF}.

Indonesia is the largest and most populous region in Southeast Asia, and its population is extremely active on social media \cite{lipman_2015}. \revision{In 2019}, there have been three critical transfers of online to offline activity: the post-elections riot, the student protest against \revision{the alleged} racist treatment of Papuan students in Surabaya, and the Jakarta student riots against \revision{the} antigraft law \cite{lamb_2019, widianto_potkin_2019, 10.2307/26938889}. 
The country also harbors local political \cite{10.2307/resrep26920.8} and religious tensions \cite{10.2307/27913255}, putting it at risk from deviant coordinated activities in social media spilling over to real-world violence.

Against this backdrop, we explore a methodology to profile coordinated activity on social media. 
We identify hashtag hijacking as a sign of coordinated activity, spotting anomalous hashtags that do not fit the conversation to discover evidence of a coordinated activity campaign and identify \revision{the} actors involved.
Hashtag hijacking occurs when a group of agents use\revision{s} a trending hashtag to promote a different message, often through adding additional unrelated hashtags \cite{vandam2016detecting}.
While this phenomenon has been studied in relation to the United States Obamacare act \cite{hadgu2013political} and troll campaigns \cite{vandam2016detecting}, it has not been sufficiently exploited as a tool to spot anomalous behavior in information manipulation campaign.
Hashtag hijacking can result in the polarization of hashtags \cite{10.1145/2487788.2487809} and conversations \cite{jackson2015hijacking}.
Next we use the BEND framework \cite{BEND}, an information manipulation framework, to characterize the actions of the actors, followed by analyzing the polarizing impact of the coordinated activity campaign.
We apply this pipeline to two contentious issues in Indonesia, one international issue involving the Palestine-Israel conflict and one regional issue involving alcoholic beverages. Our results show evidence of coordinated activity manipulating social media conversations in both events and causing polarization in the online discourse.

By providing an end-to-end characterization of coordinated activity through detection, strategy analysis and impact assessment in Indonesian social media, we aim to bridge the gap of knowledge in online coordinated activity landscape in the Southeast Asian region.

\section{Related Work}
Social media messaging is as effective, if not more, than real world persuasion in influencing opinions \cite{Bond2012}. Hence, attempts by groups of agents to manipulate online narratives are threats to society \revision{that we} should examined. 
Past work have described case studies \revision{that demonstrate these kind of coordinated agent action:} the intervention of Russia's Internet Research agency during the \#BlackLivesMatter movement \cite{Arif2018ActingTP} and the Venezuelan campaign in the 2016 US elections \cite{doi:10.1126/sciadv.abb5824}.

The core of techniques for coordinated activity detection typically rely on on the discovery of high levels of actors performing common actions within a short time window, e.g. tweet\revision{s} with the same hashtag or the same URL \cite{DBLP:journals/corr/abs-2105-07454,vargas2020detection}. 
Other methods make use of user similarity measures that combines several common actions to characterize the presence of coordinated activity on social media \cite{Nizzoli_Tardelli_Avvenuti_Cresci_Tesconi_2021}. 

Following detection of activity comes characterizing the group behavior and interaction with other agents in the social network. Several frameworks have been developed to characterize the actions and strategies in a coordinated action campaign: the BEND framework addresses communication objectives and tactics using 16 network and narrative maneuvers, four maneuvers for each letter \cite{BEND}; the ABC(D) framework describes the Actors, Behaviors, Content and Distribution \cite{alaphilippe_2020}; the SCOTCH (Source, Channel, Objective, Target, Composition, Hook) framework attempts to address the big picture by summarizing a single action to an overall campaign\cite{blazek_2021}. However, these frameworks focus on specifically the agent maneuver strategies in a coordinated activity campaign and do not account for the entire nature of the campaign, from discovery to impact.

Finally, to measure the impact and efficacy of these coordinated activities, methods range from measuring cross-community adoption of ideas \cite{Nimmo2020} to measuring development of polarization on social networks across time \cite{lee2016impact}. 

\subsection{Contributions} 
Our work combines previous approaches \revision{to characterize} online coordinated activities into a structured, holistic framework in the form of an analytical pipeline. 
We then apply the pipeline to shed light on \revision{and} illuminate the presence of coordinated activity and its effect in Southeast Asia discourse.
We study coordinated activity in two contentious issues in Indonesia, a region with political and religious tensions. 

\revision{By} using network and narrative analysis techniques, we developed a pipeline to discover, analyze and evaluate coordinated activity on Twitter. 
We apply this pipeline on two case studies in the Indonesia social media, quantitatively and qualitatively profiling one international and one regional issue. 
The pipeline begins with the detection of hashtag hijacking to discover evidence of a coordinated activity campaign and identify actors, before using the BEND framework \cite{BEND} to characterize the actions of the actors, then further analyze their in terms of polarization metrics. 
Our results show coordinated activity manipulating social media conversations are present in both events and cause polarization.
 
\section{Data Collection and Processing}
\revision{This work studies} one international and one regional issue from Indonesian Twitter discourse. We collected data with the Twitter V2 REST API on both datasets. \revision{The dataset contains only} public tweets and \revision{we do not involve any} personally identifiable data in the analysis.

For the first dataset, we focused on the impact of an international event \revision{in} Indonesian Twitter discourse. The Palestine-Israel conflict is a fundamental part of Islamic identity politics in the modern Indonesian political scene \cite{nugroho_2021}. The strong-voiced support for Palestine led Indonesia to formally call for international support to resolve the conflict in mid-May 2021 \cite{nur_2021}.
Tweets were collected from 14-20 May 2021 using the search term ``Palestina", the Indonesian word for Palestine and are filtered for tweets in the Indonesian language. This week begun with 850 rockets launched into Israel territory and ended with a ceasefire brokered by the US. In total, \revision{we collected} more than 700k tweets. 

For the second dataset, we focused on a regional event, the alcoholic beverage issue. The Indonesian president issued a decree in February 2021, which allowed several cities to receive investment in the previously off-limits alcohol industry, sparking public debate \cite{tempo_revoke}.
We collected tweets with hashtags related to the Indonesian word for alcoholic beverages ``miras" from 26 Feb to 3 May 2021. These hashtags are: \#BatalkanPerpresMiras, \#PapuaTolakInvestasiMiras and \#MirasPangkalSejutaMaksiat.
These hashtags emerged with the publication of the alcoholic beverage investment policy and faded out shortly after its revocation. 
This regional issue resulted in a smaller dataset with only 83,334 tweets from 16,991 unique agents.

Filtering for tweets in the Bahasa Indonesian language severely reduces the dataset size in comparison to datasets of English tweets due to the distribution of language in Twitter -- English is used almost ten times more often than Indonesian \cite{Hong2011LanguageMI}.

We performed bot-probability annotation using the BotHunter algorithm \cite{beskow2018bot} which has a reported 90\% accuracy. BotHunter has previously been successfully applied in Southeast Asia information studies \cite{Uyheng2021,lynnkashmir}. It extracts agent-level metadata and classifies agents using a supervised random forest method through a multi-tiered approach, each tier making use of more features. For each user agent, BotHunter provided a probability that the account is a bot.
We used a threshold of 50\%, where a probability above 50\% indicates the agent is a bot.

\section{Methodology}
\begin{figure*}[!tbp]
  \centering
  \includegraphics[width=1.0\linewidth]{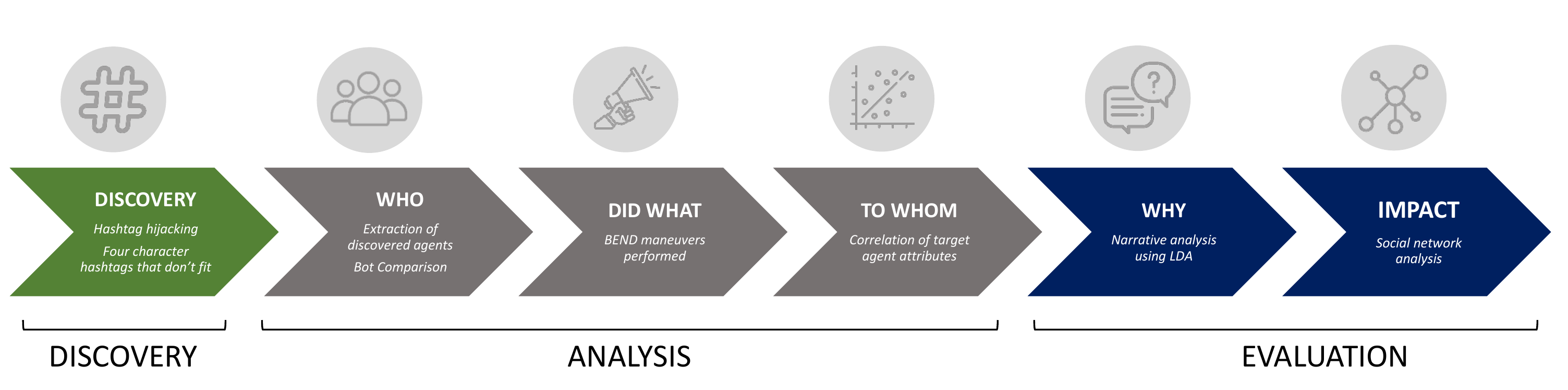}
  \caption{Methodology for discovery, analysis and evaluation of coordinated activity}
  \label{fig:pipeline} 
\end{figure*}

We detail our pipeline for discovery, analysis and evaluation of coordinated actions in the Palestine-Israel conflict and alcoholic beverages datasets. A visual summary is presented in Figure \ref{fig:pipeline}. 
All analyses were done in the Indonesian language. A native Indonesian speaker verified and translated the analysis into English.

\subsection{Discovery}

\textbf{DISCOVERY of coordinated activity.}
We define an initial social communication network using the first three days worth of data within our collection period.  In this communication network, nodes are Twitter agents. A link between two agents represents that there existed a communication between them, i.e. through retweet, @-mention etc. During these three days, we observe a surge in activity related to the hashtags used as collection parameters.

Using this initial network, we identify patterns of hashtag hijacking through anomalous hashtags. In hashtag hijacking, actors steer conversations to tangential topics by piggybacking on trending hashtags \cite{10.1145/2908131.2908179}.

We construct a hashtag-hastag co-occurence network, where two hashtags are linked together if they occur in the same tweet. 
We use the Louvain clustering technique \cite{traag_waltman_eck_2019} to discover internal groupings of hashtags in the network. This technique is a modularity-based community detection method which detects clusters with high connectivity through local optimization and aggregation \cite{Yoshida2021}. 

In both case studies, Louvain clustering results in two main types of hashtag clusters: proper-word hashtags and 4-character hashtags. We identify the 4-character hashtags as hashtags that do not fit the discourse as they are made up of four seemingly random characters. This curious hashtag group is likely an indicator of coordinated behavior.

We then identify the Twitter agents that engage in coordinated activity through hashtag hijacking within this initial network. We elaborate on this step in the next section.

\subsection{Analysis}

\textbf{WHO.}
\revision{We extract the corresponding agents and their tweets from the clusters of 4-character hashtags that do not fit.} We correlate these agents against their bot classification and observe their activity in the subsequent steps.

\textbf{DID WHAT.} 
We use the BEND framework to empirically characterize the activity of the coordinating agents. The BEND framework argues that online campaigns are comprised of sets of narrative and network maneuvers carried out by a group of actors engaging others in the social environment with the intent of altering topic-oriented communities \cite{BEND}.
A maneuver is a deliberate action performed by a social media user to achieve a desired end state. The two key types of maneuvers are narrative and network maneuvers. Narrative maneuvers make use of the messaging techniques to influence the target audience. Network maneuvers attempt to alter the structure of the network through referencing or mentioning other users within the tweets.
It has been employed in previous studies in identifying information operations activity during vaccine rollout \cite{ad21ea79e03114bf95665230d94a4b99}.
BEND \revision{detection and measurement} is provided through the ORA software \footnote{http://www.casos.cs.cmu.edu/projects/ora/software.php}, which we use to characterize the activity of the coordinating agents into 16 maneuvers. \revision{This characterization is done through a weighted average of linguistic cues derived from the tweet texts and network cues derived from the surrounding social network. The linguistic cues are calculated with a companion software NetMapper\footnote{https://netanomics.com/netmapper-government-commercial-version/}, which provides quantitative counts of psycholinguistic cues from a multi-lingual lexicon such as the number of encouragement, anger, inclusive terms. The output of these counts are used as input together with the tweet network information obtained from the Twitter data into the ORA software (v.3.0.135) for calculation of BEND scores. The ORA software calculates a score for each maneuver per agent. In our analysis, we use the mean scores for each maneuver of the agents examined at each step.}

In this study, we focus on the most prominent network and narrative maneuvers that occur in both datasets: the B- and D- maneuvers.
The B- maneuvers are four network maneuvers that provide positive manipulation of the social network, consisting of \textit{Back, Build, Boost, and Bridge}. Backing involves increasing \revision{an opinion leader's importance and effective positioning} surrounding a particular topic. For example, creative positive messaging about an individual can lead to increased social connections or followers. Build, boost, and bridge are methods that increase the social of a social community surrounding a specific topic. Building creates a group; boosting increases the size of the group; and bridging combines groups.

\revision{All the B-maneuvers are concerned with enhancing the connections among actors in the community. The Back maneuver assesses how messages support an actor or an actor's narrative, among others, by increasing their followers or likes. It considers terms that signal encouragement together with the presence of an @-mention/reference to an opinion leader, an agent with a high follower count, providing a measure of backing an influential leader in the group. The Build maneuvers are about creating a group: the measurement assesses the extent to which the messages co-mention agents, encourage agents to join, and introduce other agents. These are determined through the extensiveness of @-mentions or references to other agents, which is crucial for community building. The Bridge maneuver is about building links between existing groups so it assesses the extent to which new actors join two or more groups, introduce members of one group to another, or shares hashtags or information between groups. The Boost maneuver concerns itself with increasing the size and density of a community. It assesses the extent to which a single message mentions many actors, encourages joint activity, builds links to multiple community members, and so forth. The agent communities are determined by either co-clustering on both shared hashtags and interaction, or based on other factors of interest to the researcher. In this paper, we defined communities using network clustering methods.}

The D- maneuvers are four narrative maneuvers that negatively manipulate the message content, consisting of \textit{Dismay, Dismiss, Distort, and Distract}. Dismay messages evoke negative emotions such as sadness and anger. Dismissing disregards a topic as unimportant; distorting manipulates the primary message of the topic; and distracting changes the narrative focus to an unrelated topic.

\textcolor{teal}{
The D- maneuvers are four narrative maneuvers that negatively manipulate the message content, consisting of \textit{Dismay, Dismiss, Distort, and Distract}. Dismay messages evoke negative emotions such as sadness and anger. Dismissing disregards a topic as unimportant; distorting manipulates the primary message of the topic; distracting changes the narrative focus to an unrelated topic.
}

\revision{The D maneuvers are designed to alter who is talking about what and/or how they are talking about it, typically by shifting the conversation away or negatively affecting a view on an issue. The Dismay maneuver is used to create anger, dismay, fear or some negative emotion in the reader; consequently, the associated metric measures the presence of negative oriented sentiment, emojis or emoticons. The Dismiss maneuver is about reducing the perceived importance of a topic so it measures the extent to which there are belittling comments, statements that the topic is irrelevant, or unimportant.  The Distort maneuver is about altering what is known about an issue by introducing doubt, misinformation, disinformation or irrelevant facts; thus use of questions, equivocal terms. The Distract maneuver is about changing the topic; thus this measure considers whether a new topic is introduced, new hashtags, distance from original message, use of strong rhetorical statements, and so forth.}

\textbf{TO WHOM.}
We used the same method as in the DID WHAT section and obtain the B- and D-maneuver ratios of the posts of agents who were targeted by the coordinated activity. We match these agents with the obtained bot scores.
We then perform a correlation between the BEND scores and agent meta-data: whether the account is a verified account, the bot probability score of the account, the number of followers the account has, the number of accounts the agent is following and several network centrality values. 
The network centrality values we examined are: betweenness centrality which represents agents that lie on paths of other agents the graph; eigenvector centrality which represents agents which are connected to influential agents; and total-degree centrality which represents the agents with the most connections.
Bot and correlation analysis provides an idea of what kind of accounts are being targeted in the maneuvers. 

\subsection{Evaluation}

\textbf{WHY.}
To infer reasons behind the maneuvers, we analyze the tweet narratives of the identified coordinating agents.
We first preprocess the tweet texts by removing URLs, hashtags, @-mentions and Indonesian stopwords. 
Then we perform Latent Dirichlet Allocation (LDA) using the sklearn library \footnote{https://scikit-learn.org} on the processed tweet text in the native Indonesian language. \revision{We derived five narrative clusters through analysis of coherence scores and manual inspection.}
A native Indonesian speaker manually interpreted the word clusters. We also present the English translation alongside the original tweets in this paper.

\textbf{IMPACT.}
We assess the impact of the coordinated activity through the social network at the end of the collection timeframe. We qualitatively observe how the activity of these agents create polarized conversations through social network visualizations. 
We construct a communication network with data at the end of our collection timeframe and observe distinct subgroups. The nodes of this network represent the agents present at the end of the collection timeframe. Two agents are linked together if they had communicated with each other, i.e. via retweets or @-mentions. 
We view polarization \revision{as} the absence of interaction between distinct subgroups \cite{PRAET2021100154}. To do so, we extract the observed subgroups and quantitatively compare the Krackhardt's E/I index \cite{krackhardt1988informal} of the subgroups against the original network before agent activity. The E/I index (Equation \ref{eq:ei}) captures the extent to which a subgroup engages in actions with agents outside their subgroup relative to members within their subgroup. The index is a ratio of internal and external links and ranges from -1.0 to 1.0. A highly negative E/I index reflects the subgroup communicate within their group more than outside their group, and multiple subgroups with highly negative E/I indexes reflects polarization.

\begin{equation}
\begin{array}{l}
    \text{E/I index} = (EL - IL)/ (EL + IL) , \\ 
    \text{where EL = number of external links;} \\ 
    \text{IL = number of internal links}
\end{array}
    \label{eq:ei}
\end{equation}

\section{Results}
In this section, we detail the results of the discovery, analysis and evaluation of coordinated actions in an international and a regional event in Indonesian Twitter discourse.

\subsection{Discovery}

\begin{figure*}[!tbp]
    \centering
  \subfloat[International case study: Palestine-Israel Conflict\label{hashtaghijack-palestine}]{%
        \includegraphics[width=0.5\linewidth]{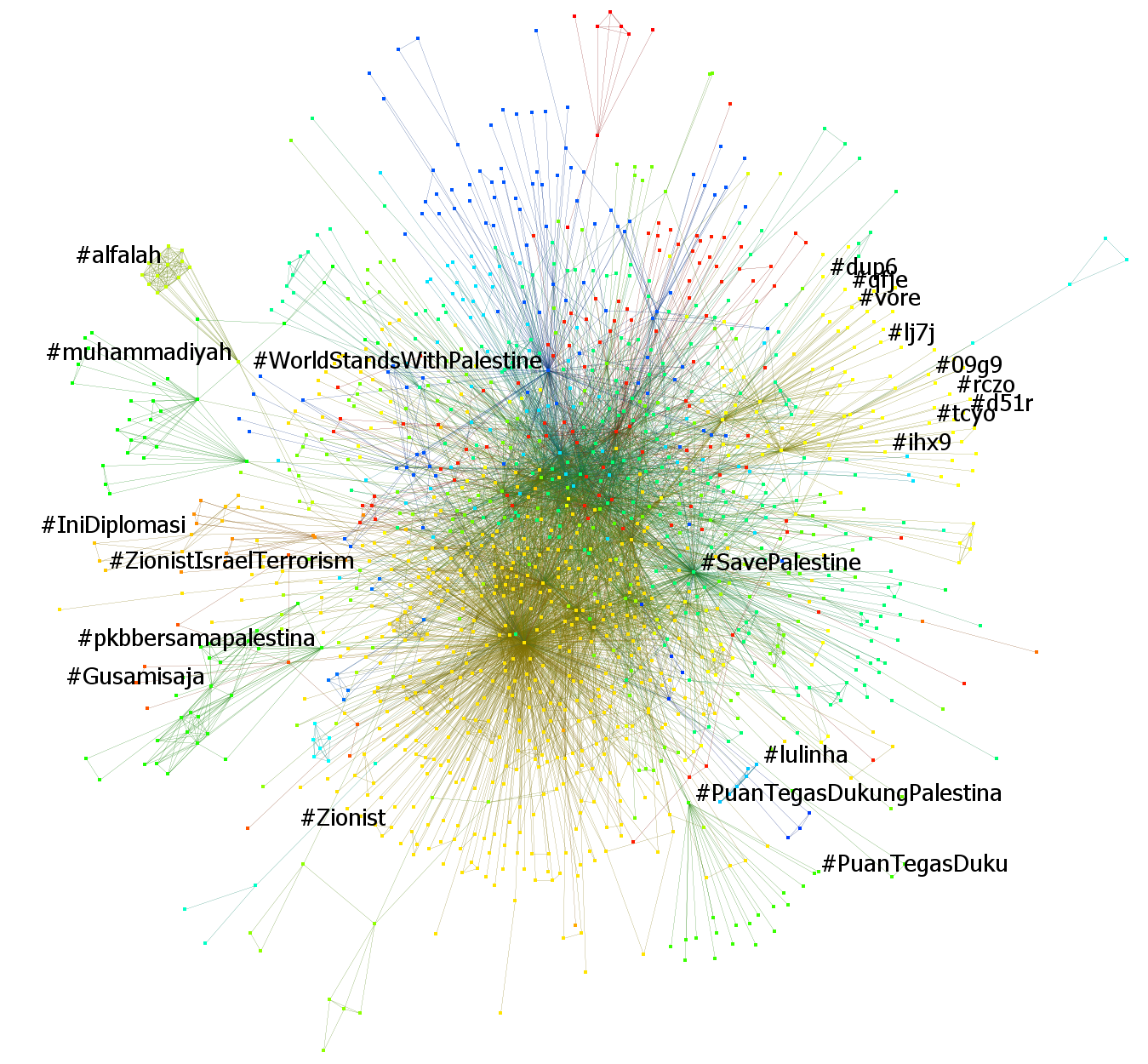}}
  \subfloat[Regional case study: Alcohol Beverages\label{hashtaghijack-alcohol}]{%
       \includegraphics[width=0.5\linewidth]{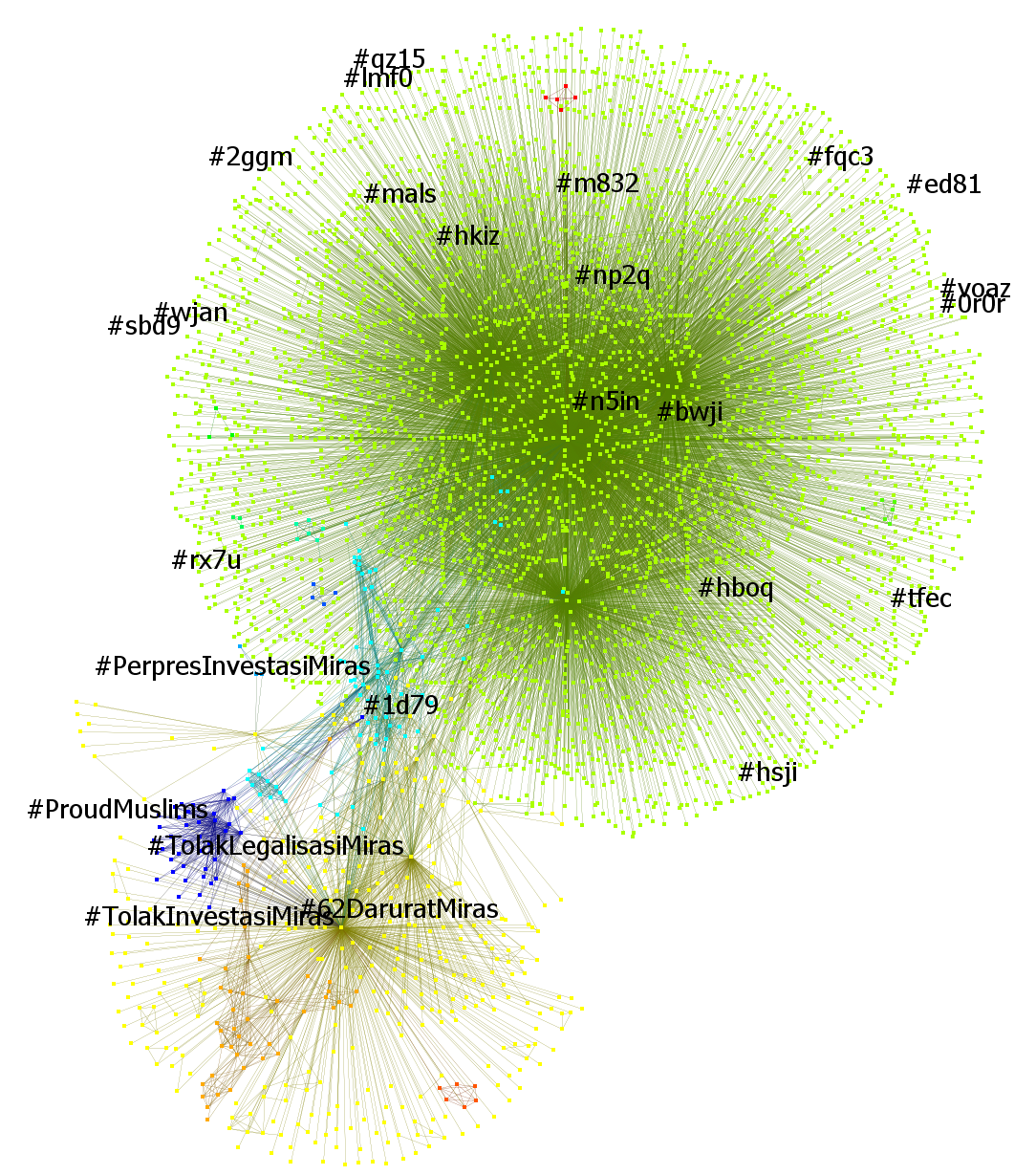}}
    \hfill
  \caption{Discovery of coordinated activity through hashtag hijacking where 4-character hashtag clusters (e.g. \#dup6, \#d51r, \#0r0r, \#qz15) co-occur with proper-noun hashtag clusters. Nodes are sized by the frequency of the hashtag usage.}
  \label{fig:hashtaghijacking} 
\end{figure*}

\textbf{DISCOVERY.} 
We discovered coordinated activity through curious clusters of 4-character hashtags (e.g. \#dup6, \#d51r, \#0r0r, \#qz15) forming around proper-word hashtags in a co-occurrence hashtag network visualized in Figure \ref{fig:hashtaghijacking}.  
We \revision{label} the agents that are using abnormal 4-character hashtags as discovered coordinated agents. We present the original and extracted dataset statistics in Table \ref{tab:hashtagdatasetstatistics}.

\begin{table}[]
\centering
\begin{tabular}{|p{2cm}|c|c|p{1.5cm}|}
\hline
\textbf{Dataset} & \textbf{Num Agents} & \textbf{Num Tweets} & \textbf{Bot \newline Percentage (\%)}\\ \hline 
\multicolumn{4}{|l|}{International case study: Palestine-Israel Conflict dataset}\\ \hline
Full & 133,728 & 714,792 & 22.08 \\ \hline 
Coordinated agents & 870 & 2,700 & 23.22 \\ \hline 
\multicolumn{4}{|l|}{Regional case study: Alcohol Beverages dataset} \\ \hline
Full & 16,991 & 83,334 & 21.12 \\ \hline 
Coordinated agents & 1,106 &  4,081 & 23.99 \\ \hline
\end{tabular}
\caption{Statistics of full dataset vs discovered coordinated agents (agents that tweet with atypical 4-character hashtags)}
\label{tab:hashtagdatasetstatistics}
\vspace{-0.7cm}
\end{table}

\subsection{Analysis}

\textbf{WHO.}
We discover 4,081 and 2,522 agents engaging in coordinated activity for the Palestine-Israel and alcohol beverages datasets, respectively. Around 23\% of each of the discovered agent groups are bots. 
Additionally, we observe that 2\% of the agents (49 agents) that first appeared in the alcohol beverages event in February 2021 emerge again in the Palestine-Israel conflict in May 2021.
These agents were all classified as bots. 

\begin{figure}[!tbp]
    \centering
   \includegraphics[width=1.0\linewidth]{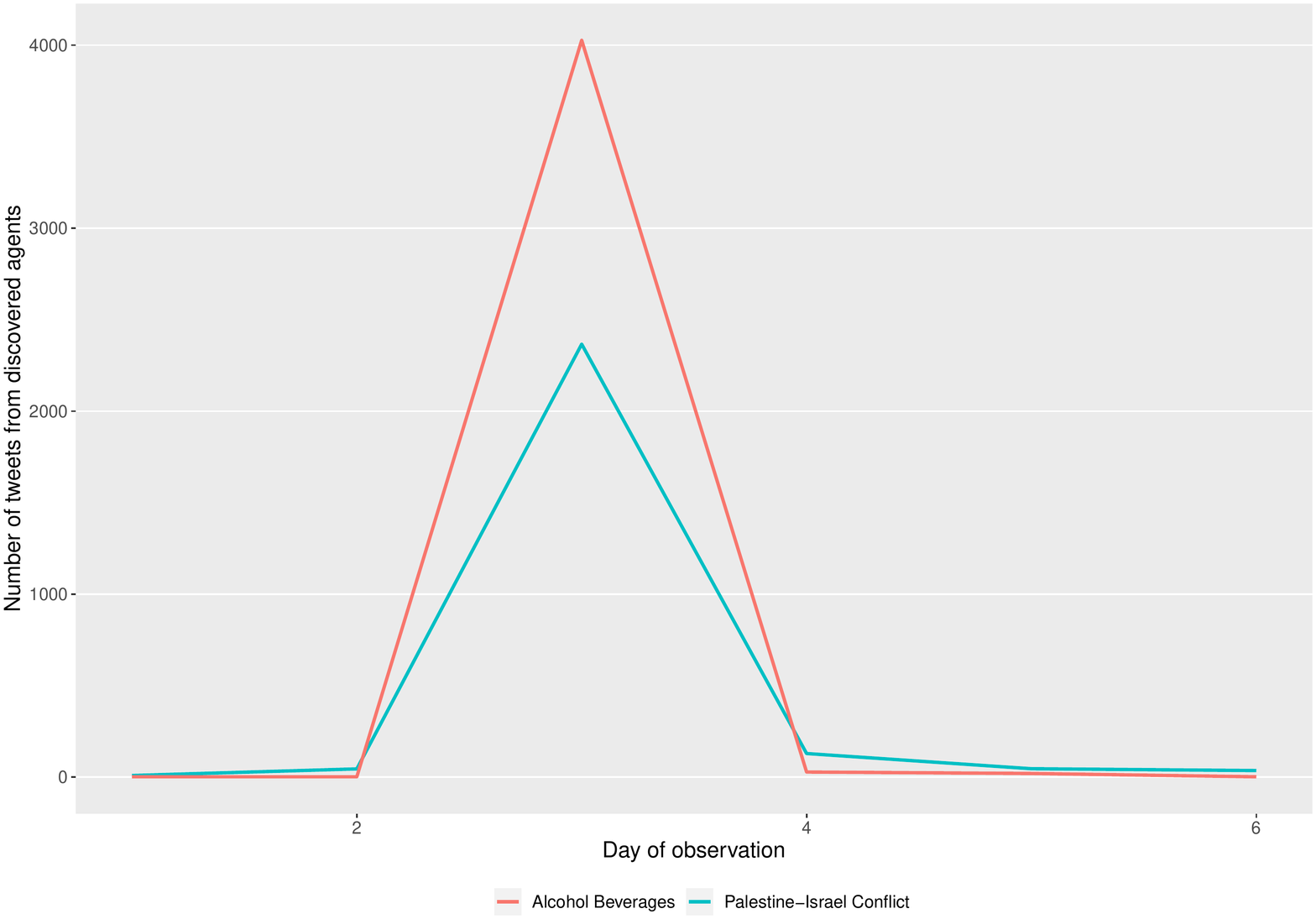}
  \caption{Tweet frequency over days of analysis for discovered agents. The coordinated agents make most of their tweets on one specific day.}
  \label{fig:graphovertime} 
\end{figure}

\textbf{DID WHAT.}
Analyzing the tweeting behavior of the coordinated agents (Figure \ref{fig:graphovertime}), we observe that the agents make most of their tweets on a short burst of time, the second or third day of our collection timeframe, suggesting synchronized tweeting behavior.

The tweets originating from these agents are mostly canned messaging that shows sympathy to the notion of Khilafah (an Islamic state in Indonesia), both independently or in conjunction with the events. The messaging templates take the templates like the following example: 

\begin{quote}
``$<$Measure of Time$>$, $<$Alhamdulillah/MasyaAllah$>$ saya melihat di kota Jakarta sudah banyak orang yang sadar khilafah, kamu gimana $<@$mention Twitter target account$>$".

English translation: ``$<$Alhamdulillah /MasyaAllah$>$ There's a lot of people in Jakarta this last $<$Measure of time$>$ that is aware of Khilafah. How about you?$<$mention Twitter target account$>$"
\end{quote}

\begin{figure}[!tbp]
    \centering
   \includegraphics[width=1.0\linewidth]{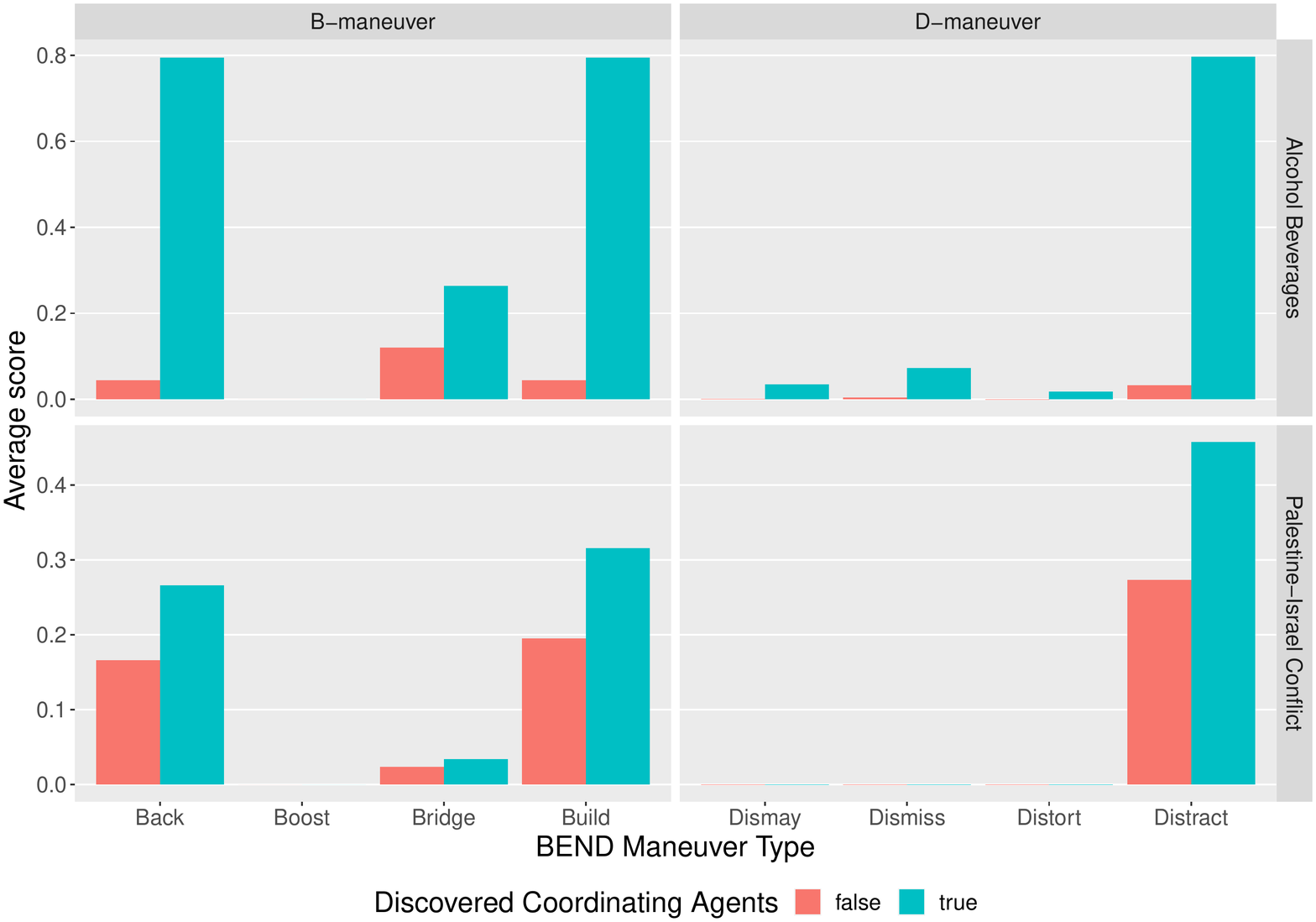}
  \caption{Average BEND score of discovered agents compared against the other agents in the dataset}
  \label{fig:bendauthor} 
\end{figure}

When \revision{we compare} the average BEND score of the discovered agents against all the other agents in the dataset (Figure \ref{fig:bendauthor}), we observe that the discovered agents score higher than the other agents, reflecting a seemingly deliberately use of information maneuvers.

The top four maneuvers the discovered agents performed are: Back (increase importance of influencers), Bridge (connect groups), Build (create groups) and Distract (post on a different topic).  
In using the Back and Bridge maneuvers, the discovered agents @-mention known influencers, which serves two purposes: first, it increases the importance of the influencer within the network; second it bridges the groups that the agent and the influencers have sway over.
When performing the Build maneuver, agents @-mention each other using a messaging template, creating an illusion of an active conversation within the group. 
The Distract maneuver supports their use of hashtag hijacking, to latch onto a trending hashtag in order to get their message across.
We present example tweets from both case studies along with their observed BEND maneuver in Table \ref{tab:maneuvrexamples}.

We did not observe Boost maneuvers because the agents were only active for short periods of time, which is insufficient for agents to grow a network. 
Another maneuver we did not observe was Dismay, or posting negative emotion, which we postulate it due to insufficient linguistic analysis libraries in the Indonesian dialect.

\begin{table}[]
\centering
\begin{tabular}{|p{4cm}|p{7cm}|}
\hline
\textbf{BEND maneuver} & \textbf{Example Tweet} \\ \hline
\multicolumn{2}{|l|}{\textbf{Case Study: Alcoholic Beverage}} \\ \hline 
Build \newline \textit{Create group or an appearance of group} & RT @I***25: Jihad dan Khilafah, Solusi Hakiki Palestina [...] \#2z22 \newline Translation: Jihad and Khilafah is Palestine's one true solution \\ \hline 
Back \newline \textit{Increasing importance of influencer} & RT @Gu***li: Klau mau jadi politisi intelektuil populis di Indonesia harusnya bela Palestina sekaligus Hamas. \newline Translation: if you want to be a populist and a politician with intellect you should defend palestine and hamas \\ \hline 
Bridge \newline \textit{Connect groups to audience of known influencers, increasing importance of that influencer} & @Su***20 @Ros***HQ @Yua****sna @Yo***be (targeted influencers)  \textit{Meski keadaan ekonomi negeri ini defisit, tidak lantas hrs melegalkan sesuatu yang ALLOH SWT haramkan} \#MirasIndukMaksiat \textbf{\#zyl7} 
\newline Translation: Despite the economic deficit, doesn't mean we need to legalize something God forbid
\\ \hline 
Dismiss \newline \textit{Post on unimportance of topic} & \textit{Negeri-negeri umat muslim hanya mampu mengutuk atas perbuatan israel tapi tidak mampu membebaskan palestina} \#AqsaCallsArmies \textbf{\#m4p8} \newline Translation: Countries with muslim population can only condemn but are not able to free Palestine
 \\ \hline 
Distort \newline \textit{Alter main message of topic} & \textit{Sejak tidak ada khilafah. Palestina mejadi negri terjajah} \#AqsaCallsArmies \textbf{\#2nw9} \newline Translation: Since Khilafah is gone, Palestine is a colonized nation
\\ \hline 
Distract \newline \textit{Post on different topic} & \textit{Sebab ketidakadilan yg diterapkan kapitalisme sekuler, membuat israel jumawa dan semakin berkuasa atas saudara di Palestina} \#AqsaCallsArmies \textbf{\#892t} \newline Translation: Because of the injustice brought forth by secular capitalism, Israel is free to usurp power and exert arrogance over Palestinians
\\ \hline 
\end{tabular}
\caption{Examples of Tweets for each maneuver. The 4-character hashtag used to discover the coordinated activity is highlighted in bold text. User accounts that are @-mentioned are redacted to preserve user privacy.}
\label{tab:maneuvrexamples}
\end{table}

\begin{figure*}[!tbp]
    \centering
   \includegraphics[width=1.0\linewidth]{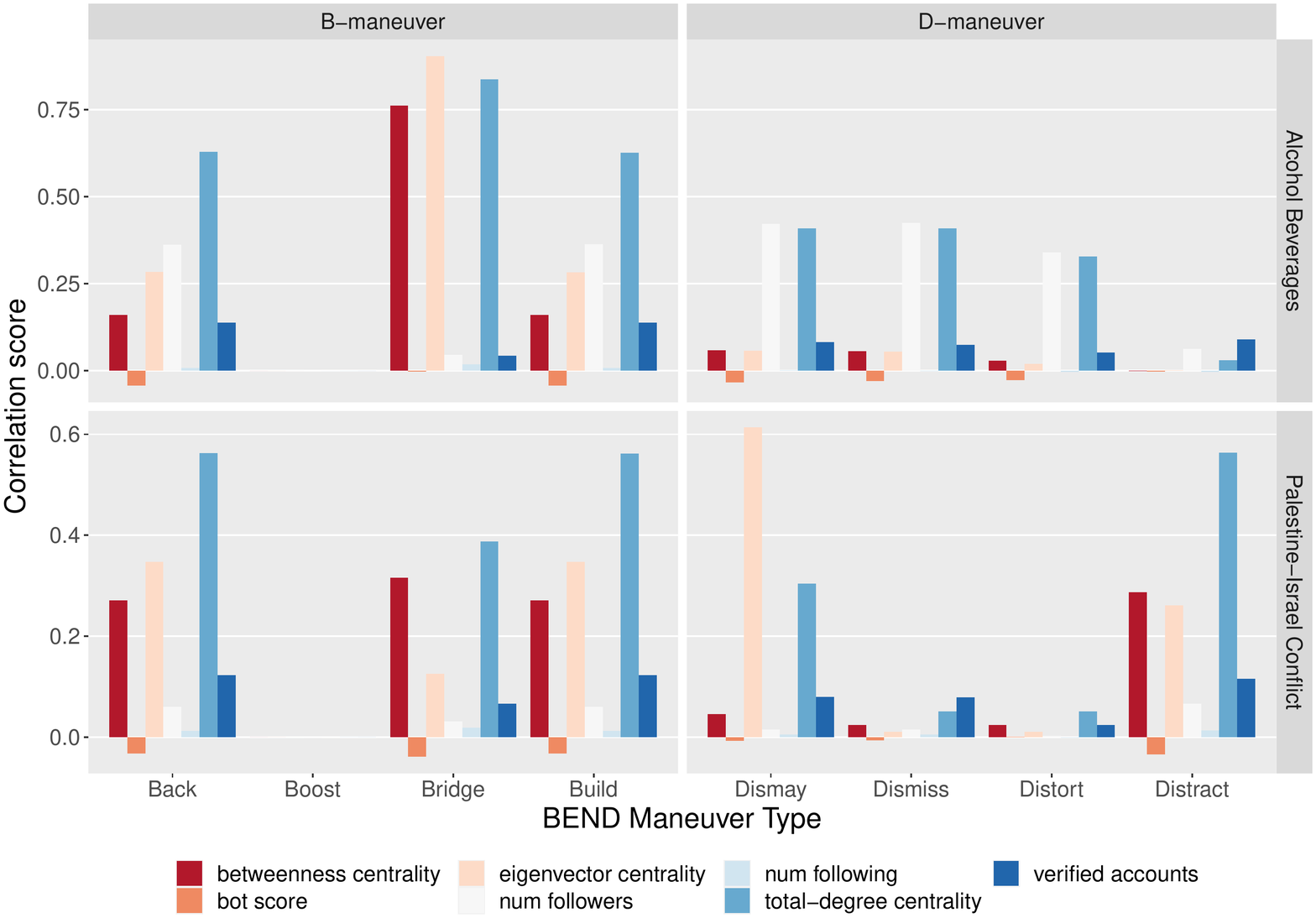}
  \caption{Correlation scores of TO WHOM the BEND maneuvers were addressed.}
  \label{fig:towhomcombined} 
\end{figure*}

\textbf{TO WHOM.}
We obtained the B- and D-maneuver ratios of agents that were targeted by the coordinated activity and performed a correlation analysis of the \revision{ratios between the BEND maneuver scores agent meta-data scores. These ratios are referred to as ``correlation scores'' and plotted in Figure \ref{fig:towhomcombined}.}

Across both international and regional case studies, we observed positive correlation of maneuver ratio with centrality values, indicating that maneuvers were performed \revision{influential agents} in the network. All maneuvers were employed on agents with high total-degree centrality and large number of followers, indicating the leveraging on the target agents' connections. There is a lower positive correlation between maneuver and the number of verified accounts, suggesting that maneuvers were performed on both verified and non-verified accounts. Finally, we observed negative correlation with bot probability score, signifying that maneuvers were more often performed on non-bot accounts.

\subsection{Evaluation}

\textbf{WHY.}
Through narrative analysis using LDA, we observe the formation of two main narrative themes for each case study, indicating the deliberate direction of information narratives towards those topics. 
The themes, its derived sampling of native keywords and their English interpretation are shown in Table \ref{tab:narrativework}.
In the Palestine-Israel conflict study, The themes show a division between a diplomatic view to a radical and militaristic view.
In the alcoholic beverages study, the themes show an emerging polar shift from discursive opinions to reactionary calls for actions.

\begin{table*}[]
\begin{tabular}{|p{12cm}|}
\hline
\textbf{International Case Study: Palestine-Israel Conflict} \\ \hline 
\textbf{Narrative Theme A:} Strong but diplomatic support \\ \hline 
palestina israel indonesia yahudi saudara perang hamas zionis orang urusan tanah donasi allah bela gaza kalau rakyat rumah mereka muslim \newline
\textit{Translation: Let's donate and pray for the houses and lands of our muslim brothers in palestine in a midst of a struggle with israeli jewish zionists } \\ \hline 
palestina israel indonesia negara orang gaza kemanusiaan kita mendukung tidak rakyat hamas membela pemerintah serangan pernyataan tanah sikap teroris negeri \newline \textit{Translation: Palestine-Israel is a humanitarian thing, don't put Hamas as a terrorist - Indonesia and their people have to make a stand} \\ \hline 
\textbf{Narrative Theme B:} Radical messaging \\ \hline 
palestina bebaskan israel masalah tentara agama bagi akidah allah solusi kemanusiaan orang bersatu kirim islam perkara umar bukan dunia penguasa \newline 
\textit{Translation: Leaders have to muster army and unite to liberate palestine - it's an islamic thing to do so} \\ \hline
palestina tanah muslim kaum israel milik warga yahudi zionis muslimin khilafah negara islam anakanak ratusan puluhan tertumpah netizen pemimpin solusinya \newline 
\textit{Translation: The solution to the landgrabs being done by israeli jewish zionists to our muslim lands in Palestine is Khilafah} \\ \hline 
\textbf{Regional Case Study: Alcoholic Beverage} \\ \hline 
\textbf{Narrative Theme C:} Clear call to action \\ \hline 
khilafah malam dakwah kota miras jakarta islam pejuang ikutan mengemban merusak akal tagar naikin trending umat generasi hashtag orang siang \newline 
\textit{Translation: Let's join the fight against mind-destroying alcoholic beverage and spread words for Khilafah by making this hashtag trending} \\ \hline
malam muslim allah miras yang tagar suaranya naikin ikutan islam syariat syariah khamar sadar perjuangan jadikanlah umat kapitalisme sistem perbuatan \newline 
\textit{Translation: Let's crank this hashtag to increase awareness against the system of capitalism that promotes alcoholic beverages and uphold syaria as per God intended} \\ \hline 
\textbf{Narrative Theme D:} Discursive Tweets \\ \hline 
miras industri investasi haram bangsa membuka tolak menghancurkan papua muslim perpres generasi allah izin buka daerah dilegalkan presiden merusak keras \newline
\textit{Translation: Opening investment in the alcohol industry per the president's rule would destroy the younger generation} \\ \hline
miras rakyat indonesia bangsa trending menolak mabuk negara saham tagar yang komen setiap islam buzzer mengkaitkan postingan pempr membuktikan manjadi  \newline \textit{Translation: This trending post proves Indonesian people is against drunkenness and alcoholic beverages despite what buzzers say} \\ \hline 
\end{tabular}
\caption{Main narrative clusters for each case study. We provide a sampling of native Indonesian keywords and an English translation. The narrative theme alphabets correspond to network clusters in Figure \ref{fig:overallimpact}. 
}
\label{tab:narrativework}
\end{table*}

\begin{figure*}[!tbp]
    \centering
  \subfloat[Impact of D-maneuvers on \textbf{International case study: Palestine-Israel Conflict}\label{fig:impact-palestine}]{%
        \includegraphics[width=0.50\linewidth]{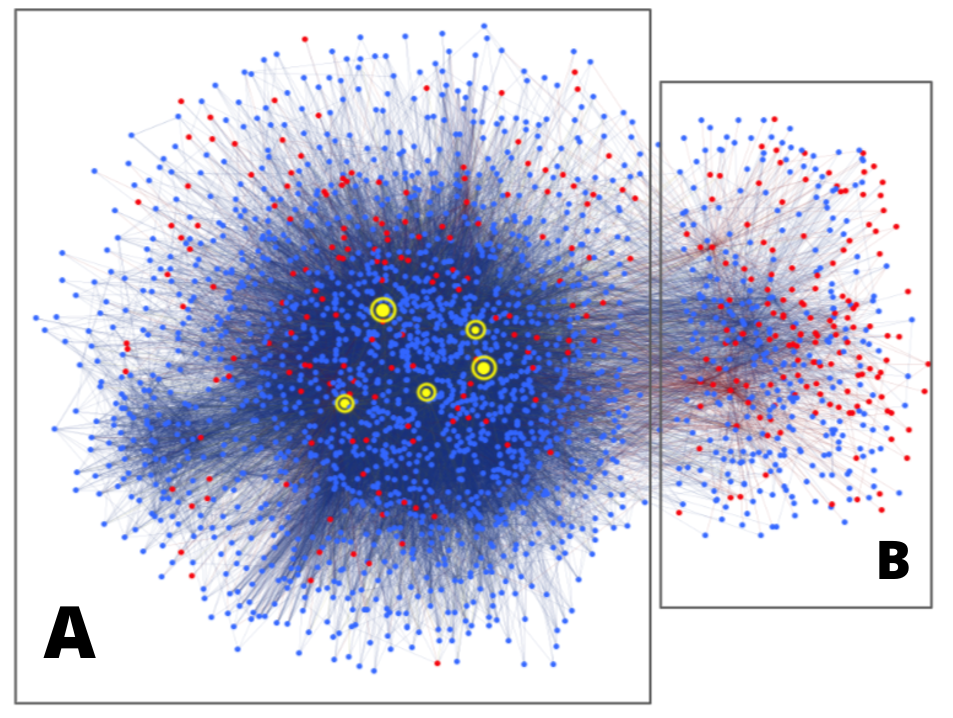}}
  \subfloat[Impact of B-maneuvers on \textbf{Regional case study: alcohol beverage}\label{fig:impact-alcohol}]{%
       \includegraphics[width=0.50\linewidth]{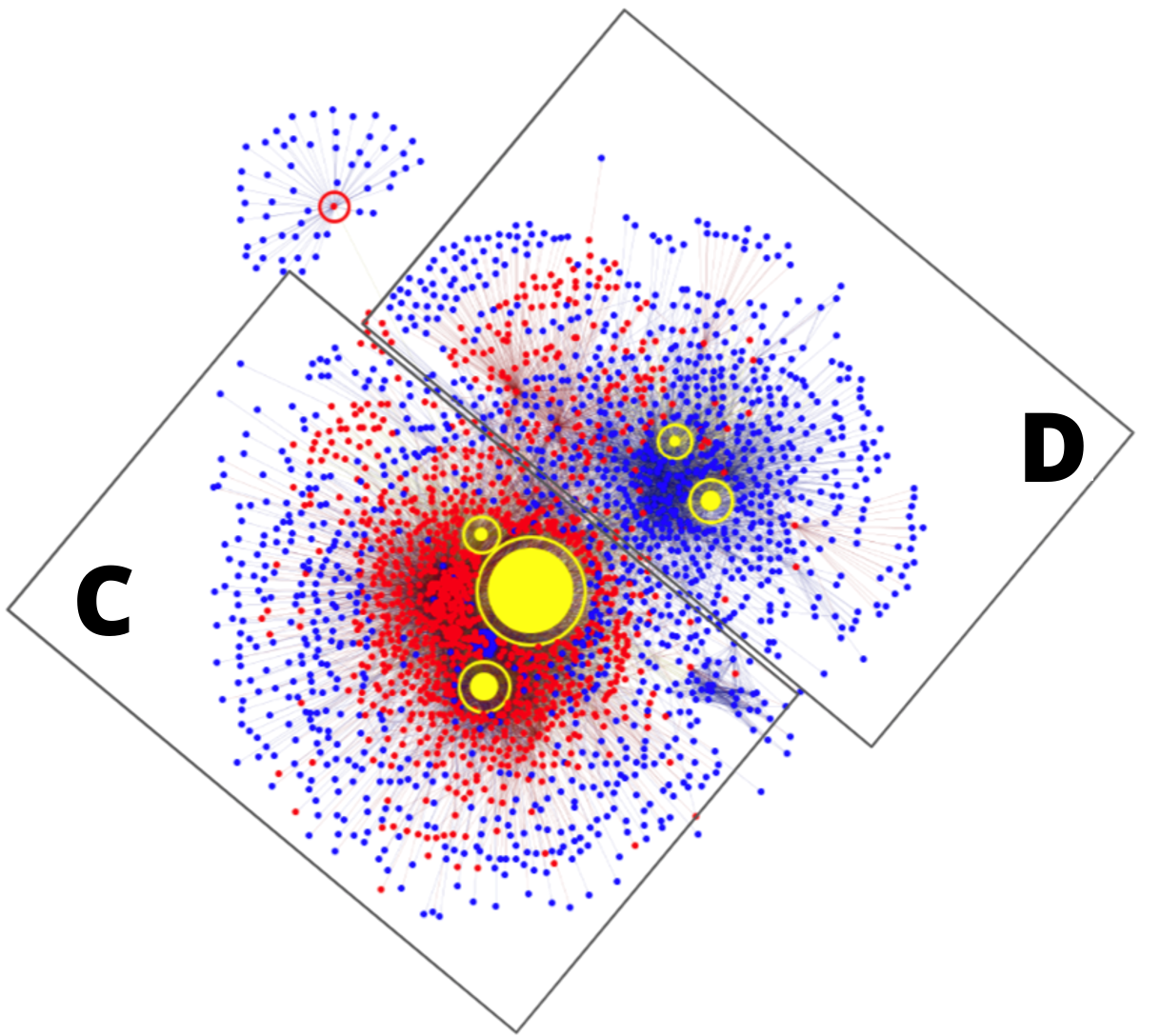}}
    \hfill
  \caption{Impact of B- and D- maneuvers on the social network. Red nodes represent the discovered agents and blue nodes are general agents in the network. Key agents in the network are highlighted in yellow and sized by their total-degree network centrality value.}
  \label{fig:overallimpact} 
\end{figure*}

\textbf{IMPACT.}
A qualitative visual inspection shows the formation of polarized networks at the end of the data collection, after the coordinated activity had taken place. This is consistent with our finding in the previous WHY section of two main narrative clusters, where the discourse occurs in clusters.
Figures \ref{fig:impact-palestine} and \ref{fig:impact-alcohol} show the impact of the D- and B-maneuvers on the Palestine-Israel and alcohol beverages network respectively, split into distinct clusters using the Louvain clustering technique.
The figure also highlights the key agents with high centrality measures in the communication network and are potentially targets of the maneuvers. Agents discovered to be part of coordinated efforts through the hashtag-hijacking method are colored in red, while other agents are colored in blue. 

We perform quantitative measures on these subgroups \revision{by evaluating} of the E/I index on the communication network of each subgroup. We present the results in Table \ref{tab:impactmetrics}. We observe a highly negative E/I index across all subgroups after the presence of coordinated activity.
This highly negative E/I index reflects the dominance of internal over external ties. Most communication links are internal links, meaning agents communicate largely within their own subgroups rather than with members from outside their group. 
This measure highlights the polarizing effect of coordinated agent activity on both case studies.

\begin{table}[]
\centering
\begin{tabular}{|c|c|}
\hline
\textbf{Narrative Theme} & \textbf{E/I Index} \\ \hline
\multicolumn{2}{|l|}{\textbf{International case study: Palestine-Israel Conflict}} \\ \hline 
Before agent activity & 0 \\ \hline 
A & -0.060 \\ \hline 
B &  -0.950 \\ \hline 
\multicolumn{2}{|l|}{\textbf{Regional case study: Alcoholic Beverage}} \\ \hline 
Before agent activity & 0.005 \\ \hline 
C &-0.660 \\ \hline 
D &  -0.370 \\ \hline
\end{tabular}
\caption{Metrics for measure of impact of discovered agent activity on the social network. After coordinated agent activity, the E/I index is highly negative, highlighting the polarizing effect of the coordinated activity.}
\label{tab:impactmetrics}
\end{table}

\section{Discussion}
\textbf{DISCOVERY.}
Despite starting off with a pretty large dataset (~700k and ~80,000 tweets), the number of discovered coordinating agents is only about 6.5\% for both datasets. 
\revision{This study shows that such a small percentage of agents can effectively polarize a conversation. This finding } is a worrying trend, indicating that online discourse does not require large agent farms to create an impact.

Within the discovered agents of both datasets, we observe that the proportion of bots are lesser than non-bot agents. While bots have been observed to perform coordinated activity in social network \cite{7395652,alassad2020developing,overbey2019using}, the composition of the discovered agents suggests human-bot teaming at play. 
These findings are consistent with past work surrounding the NATO Trident Juncture Exercise, where the messages appeared to be driven by cyborg accounts by assisted automated messaging \cite{tridentpipeline}.

Bot agents comprised of around 23\% of each dataset. 
These agents post tweets with canned messaging, following an easily identifiable template. \revision{Some agents appear in both of the cases:} they first appeared in the alcohol beverages case study in February 2021 then die\revision{d} down before emerging again in the Palestine-Israel conflict case study in May 2021. In the regional alcohol beverages conflict, they performed the B-maneuvers to support groups criticizing the legalization of alcohol beverages. In contrast, they performed the D-maneuvers \revision{in the international conflict} to distort the Muslim population's view of the conflict. This evolution of bot strategy suggests possible tactical planning to the actions, designed to invoke responses from the interacting agents. 

\textbf{ANALYSIS.}
Social media has been widely documented to be a hotbed for narrative propagation due to its popular usage \cite{thompson_2011}.
\revision{Using the BEND framework,} we quantified the nature of the coordinated activity and how the agents involved used narrative and network maneuvers to gain influence and spread messages to a larger audience. 

The discovered agents engaging in coordinating activity have been identified to target those in the network with strong connections to others, as evidenced by the high positive correlation between total-degree centrality and number of followers. At the same time, they also target non-bots. The combination of these two factors is cause for concern, because influential people can incite real-world violence. One example is the 6 January 2021 Capitol Hill riot which was very likely incited by a tweet from then-president Trump \cite{bbcnews_2022}.

Discovered agents in the regional case study (alcoholic beverages) targeted other agents with large number of followers. \revision{In contrast,} discovered agents in the international case study (Palestine-Israel) targeted agents with high betweenness centrality values. This could reflect the change in selection of tactics of the discovered agents: in the regional case, they selected other agents with a large number of followers to spread their message while in the international case, they targeted agents whom many other agents have a path through.

In the DID WHAT analysis, there are indications that the coordinating agents perform maneuvers on the social network to increase their group's influence and make themselves more central to the network. They also spread their core narrative through templated tweets to tie Khilafah as a solution to all societal problems, from alcohol-driven crime to the Palestine-Israel conflict.
Putting these observations together, we postulate that discovered agents are leveraging on public discourse to disseminate hard-line beliefs to a larger group through a context \revision{in which the public will be more sympathetic.}
These information operations actions created polarized discourse - which would slow down both public resolve and government response if a crisis brewed around that topic; an example happened during the COVID19 pandemic \cite{10.3389/fpos.2021.622512}.

\textbf{EVALUATION.}
In our evaluation of coordinated activity, while we cannot verify nor pinpoint specifically the exact intent of the coordinating agents, we make our inferences through their actions and the narratives they purport. 
In the Palestine-Israel conflict case study, these agents call out that a Islamic state can save Palestine. 
In the alcohol beverages case study, these agents call out for an Islamic state through messages on Khilafah philosophies as a possible solution to the chaos created by the alcoholic beverage debacle.
We infer that the public discourse is used as a pathway to reach potential sympathizers for these narratives. 
The repeated exposure provided by a long-winded online discourse could induce intrinsic psychological gain for narratives in exposed agents \cite{zhou_zhao_lu_2015}.

\revision{In this dataset, the spurt of coordinated activity only happened during a short timeframe within the data collection window. We choose the data collection window for it coincided with contentious issues in the region. This short activity burst of the agents is another indication that these online maneuvers are likely deliberate, coordinated activity.}

To characterize polarization in this study, we used the diversity of topics, the ratio of external-internal link index and visual inspection of the social network to measure polarization. With these measures, we observed two distinct groups for each case study, leading us to infer the impact of discovered agent activity on segregating the groups.

Both narrative and network analysis of the aftermath of the maneuvers show two separate groups, indicating the \revision{agents' success} in polarizing the groups. 
The observation of coordinated activity and the resulting polarization is dangerous given Indonesia is facing a trend of rising Islamic conservatism \cite{thompson_2017}. 
Additionally, with the observation that a handful of agents appeared across both events and the modus operandi of both case studies are similar, we postulate that the two coordinated action campaigns may be related.
These could set the stage for future factional conflicts that can turn to destabilizing conditions in countries with diverse communities like Indonesia, where tolerance is an essential tool in keeping the country intact \cite{menchik_2021}.

\textbf{Limitations and Future Work.}
Several limitations nuance our conclusions in this work. First, sampling Twitter data remains limited by API generalizability issues, suggesting caution in extrapolating findings. 
Second, while the discovery of coordinated actors through hashtags that do not fit (i.e. four-character hashtag) provides a systematic way of identifying an ongoing campaign, this may not be a comprehensive illustration of the discovery of coordinated campaigns. Future work should combine these hashtags with temporal information, i.e. appearance of these hashtags within a short timeframe of other co-hashtags.
Last, the workflow consists of several critical steps: extracting agents from texts, deciding whether a user is a bot, creating social networks, calculating network centrality and maneuvers metrics, then deriving conclusions. Minute changes in each step may affect the \revision{findings} of subsequent steps. While better accuracy tools may be called for, one direction of mitigation is to perform analysis across multiple timeframes and lend more weight to the analysis if the timeframes support each other. 
Future work also calls for correlating coordinated action against offline activities to access the social impact of discovered coordinated actions beyond polarizing online conversations. \revision{It also calls for looking at similar coordinated activities across discourse that spans longer periods.}

\section{Conclusion}
In this paper, we make an initial attempt to understand deviant coordinated behavior in the Indonesian Twitter discourse. To this end, we collected two datasets on controversial issues, one international and one regional. We analyze the datasets in terms discovery, analysis and evaluation of coordinated activity. 
We show that coordinated activity can be detected through hashtag hijacking and measured through network and narrative analysis techniques.
In both case studies, we observe the worrying trend that a small group of coordinated agents (around 6.5\% of agents collected) can cause narrative and network polarization. 
\revision{We hope} our study has shed light on understanding social phenomena in an understudied region of Southeast Asia. The early detection of coordinated effort online may present the possibility of stopping any resulting violence in the real world.

\backmatter

\section*{Declarations}
\subsection{Affiliations}
\textbf{Center for Computational Analysis of Social and Organizational Systems, Carnegie Mellon University, Pittsburgh, Pennsylvania, United States}

Adya Danaditya, Lynnette Hui Xian Ng and Kathleen M. Carley

\subsection{Acknowledgements} 
Adya Danaditya would like to thank Lembaga Pengelola Dana Pendidikan, the Indonesian government scholarship grantor that funded the author’s study at Carnegie Mellon University.
The views and conclusions are those of the authors and should not be interpreted as representing the official policies, either expressed or implied, of any government or organization.

\subsection{Contributions}
All authors contributed to the conception of the study. Adya Danaditya performed the data collection, analysis and writing edits. Lynnette Hui Xian Ng contributed to the analysis and writing. All authors read and approved the final manuscript.

\subsection{Data Availability} Data can be made available upon reasonable requests in accordance to the Terms of Conditions of Twitter. Please contact the authors for data requests.

\subsection{Ethical Approval}
We did not conduct any experimental research involving humans or animals.

\subsection{Conflict of interest} All authors declare that they have no conflict of interest.

\subsection{Corresponding Author}
Adya Danaditya and Lynnette Hui Xian Ng

\bibliography{sn-bibliography}


\end{document}